\documentclass[twocolumn,showpacs,preprintnumbers,aps,pra,amsmath,amssymb,floatfix]{revtex4-1}

\usepackage{graphicx}
\usepackage{dcolumn}
\usepackage{bm}
\usepackage{epsfig}
\usepackage{color}
\usepackage{natbib}

\begin{document}

\title{Production of a rubidium Bose-Einstein condensate in a hybrid trap with light induced atom desorption}

\author{Dezhi Xiong, Fudong Wang, Xiaoke Li, Ting-Fai Lam, Dajun Wang}
\email{djwang@phy.cuhk.edu.hk}
 \affiliation{
 Department of Physics and Center for Quantum Coherence, The Chinese University of Hong Kong, Shatin, New Territories, Hong Kong SAR, China }

\date{\today}

\begin{abstract}
We report on the production of a rubidium Bose-Einstein condensate in a simplified vacuum apparatus. Magneto-optical traps with large numbers and ultra-high vacuum for moderately long conservative trap lifetimes of 16 seconds are sequentially obtained with light induced rapid atomic vapor pressure modulation. Subsequent evaporative cooling is carried out in two stages in a hybrid magnetic quadrupole plus optical dipole trap. High evaporation efficiencies are observed in both stages and $^{87}$Rb BECs with more than $10^{5}$ atoms can be reliably produced with total evaporation time of only 9.5 seconds.

\end{abstract}

\maketitle


\section{Introduction}

Bose-Einstein condensates (BEC) of dilute atomic gases were first produced in 1995\cite{Anderson95,Davis95,Bradley95} using laser cooling followed by evaporative cooling. To date, these two cooling methods are still the standard and indispensable steps for ultracold quantum gas studies. Evaporative cooling prefers ultra-high vacuum (UHV) for efficient thermalization, whereas collecting large number of atoms by laser cooling needs enough particles in the background. These two contradictory requirements are the main reasons behind the complexity of typical BEC UHV setups. For instance, the Zeeman slower\cite{Phillips82} and the double magneto-optical trap (MOT) system \cite{Myatt96} are two of the most common implementations to overcome this issue. Here we describe a setup consists of a simple single glass cell. With the ultraviolet (UV) light induced atomic desorption (LIAD), we meet the vacuum conditions at both ends and are able to produce Rb BECs with more than 10$^5$ atoms repeatably.

LIAD has been used to produce Rb BECs \cite{Du04} and $^{40}$K degenerate Fermi gases \cite{Aubin06} on atomic chips in single glass cells previously. The strong confinement provided by the chip makes evaporation happen very rapidly. However, the presence of a surface only hundreds of microns from the atoms makes the application of an optical dipole trap (ODT) very challenging. The ODT is especially important for our future plan of studying Feshbach resonances\cite{Kohler06,Chin10} and ultracold polar molecules\cite{Ni08}. In a stainless chamber, Mimoun et al.\cite{Mimoun10} achieved a sodium BEC with LIAD. Evaporation was carried out in pure optical traps. To maintain efficient evaporation to reach degeneracy, besides a crossed-beam ODT, an additional tightly focused ``dimple'' beam had to be added. In the current work, we have adapted the hybrid trap developed by Lin et al. \cite{Lin09} where a spherical quadrupole trap is used for the first stage of evaporative cooling. The pre-cooled atoms are then transfered to crossed or single beam ODTs with foci displaced from the magnetic field zero for further evaporation to BEC.

Compared with the ODT, a magnetic trap typically has deeper trap depth and larger trap volume. It can be mode-matched well with laser cooled atom clouds for efficient conservative trap loading. The simplest magnetic trap is of the spherical quadrupole configuration, which can be easily generated from a pair of anti-Helmholtz coils. Due to its superior confinement\cite{Ketterle99}, evaporation in this linear trap is more efficient compared with harmonic Ioffe-Pritchard traps before Majorana loss worsens at low temperatures. The Majorana hole can be plugged with a blue detuned laser beam focused to the quadrupole trap center \cite{Davis95,Dubessy12,Heo11}. Lin et al. instead used a far red detuned laser focused below the quadrupole center to displace the overall trap potential minimum away from zero B field \cite{Lin09}. This method is easier to implement in terms of optical alignment. It also results in a BEC in ODT without other intermediate steps. With full strength quadrupole trap, this ODT does not prevent Majorana loss completely. Subsequent quadrupole trap decompression transfers a large portion of the pre-cooled atoms to the new trap center defined by the ODT. Majorana loss is eliminated this way with the added huge gain of phase-space density because of the trap shape deformation and evaporation during the transfer. In this work, we explore evaporative cooling of Rb to BEC in this hybrid trap in a very simple vacuum setup. We believe that this compact system can be useful for new groups who would like to setup a BEC apparatus quickly with relatively low cost.

\section{Experimental setup}
\subsection{Vacuum system and LIAD source}
At the center of our BEC setup is a single rectangular glass cell without anti-reflection coating. The cell has outer dimensions of 100 mm$\times$40 mm$\times$40 mm and is connected to a standard CF35 cube. Rubidium (and sodium for future experiments) dispensers (Alvatec GmbH) are directly inserted into the glass cell from the opposite side of the cube. The distance between the dispensers and the cell center is $\sim$ 12 cm, ensuring alkali atoms released from the dispensers have the most direct line-of-sight with cell walls for fast atom absorption. The vacuum is maintained by an ion pump(Gamma Vacuum 45S). A titanium sublimation pump (Varian Vacuum TSP Cartridge) is also installed, but it has never been fired after the initial vacuum preparation stage. Standard procedures are followed to obtain UHV. At the ion pump position, the pressure reads $1.8\times10^{-11}$ torr. The pressure at the center of the glass cell is estimated to be 5 to 6 times higher limited by conductance. This is consistent with the measured magnetic trap lifetime as discussed in later subsections. 

Initially, to coat the glass cell walls with Rb atoms, 2.5 A current is applied to one dispenser for a day. Later, firing the dispenser for 15 minutes once a week is enough to replenish atoms. It is then left off during further experiment cycles. The LIAD light is provided by a 365 nm UV LED (Thorlabs M365L2) with $\sim$ 200 mW uncollimated power output at its maximum current. It is mounted close to the cell to reach enough localized intensity for efficient desorption. We do observe saturation behavior of the Rb MOT number with increasing UV light power. But it is suspected that with multiple LEDs to cover different parts of the cell walls, atom numbers could be further increased. The LED current driver can be quickly modulated by a TTL signal synchronized with other events. Similar to other groups\cite{Telles10,Klempt06} using LIAD, we find that the vacuum pressure can recover within a short time scale after switching off the UV light. To allow vacuum recovery, the LED is typically switched off one second before magnetic trap loading. The MOT number loss during this interval is less than 10$\%$.

\subsection{Laser cooling}

Our MOT is in the real six-beam configuration with total laser power of 70 mW and $1/e^2$ beam diameters of 25 mm. The more convenient three beam retro-reflection configuration does not produce as good initial phase-spaced densities (PSD) in the MOT and molasses cooling stages. We use two home-built external cavity diode lasers (ECDL) \cite{Arnold98} to provide the trapping and repumping beams. The trap laser frequency is empirically detuned -19 MHz from the $F=2$ to $F'=3$ cycling transition. The repump beam has a power of 10 mW and is on resonance with the $F=1$ to $F'=2$ transition. The trap power is boosted up with a 150 mW laser diode injection locked by the trap ECDL and then delivered to the UHV cell by polarization maintaining fibers.           

With LIAD, the MOT loading time constant is typically 10 seconds. More than $2\times10^8$ Rb atoms can be collected after 20 seconds loading. The atoms then undergo a 24 ms compressed MOT stage by reducing the repump power to $\sim$ 150 $\mu$W and increasing the trap laser detuning to -32 MHz. A 6 ms polarization gradient cooling stage is then applied by abruptly turning off the 10 G/cm gradient and further detuning the trap laser to -78 MHz. The repump laser beams are then turned off first, followed by the trap beams 1 ms later. Following these steps, $>$ 95$\%$ atoms in the $F$ = 1 manifold with temperatures around 15 $\mu$K and densities $>$ 10$^{11}$/cm$^3$ are routinely achieved. After optically pumping the atoms to the $|F = 1, m_F = -1 \rangle$ state, $> $ 65$\%$ of them can be loaded into the magnetic quadrupole trap.     

\subsection{The hybrid trap}
The magnetic trap is generated from the same anti-Helmholtz coil pair used for the MOT field. They are wound with standard 4 mm outer diameter refrigerator copper tubes, in-house insulated by doubly-wrapped Kapton tapes. Pressurized water is running continuously inside to remove the resistive heating. The current applied to the coils is actively stabilized with the help of a current transducer (LEM IT200-S) and a simple electronic feedback servo. For this experiment, we use an axial field gradient of 160 G/cm. 

To capture the laser cooled atoms with minimum PSD loss, we first abruptly turn on the magnetic field gradient to 60 G/cm. After a holding time of 70 ms, it is ramped up to 160 G/cm in 130 ms. This adiabatic compression heats the cold atoms up to 90 $\mu$K, but the peak collision rate increases by a factor of 3.7. The PSD of 2.1$\times$10$^{-6}$ is an excellent starting point for evaporation.

 \begin{figure}[hbtp]
 \includegraphics[width=0.85\linewidth]{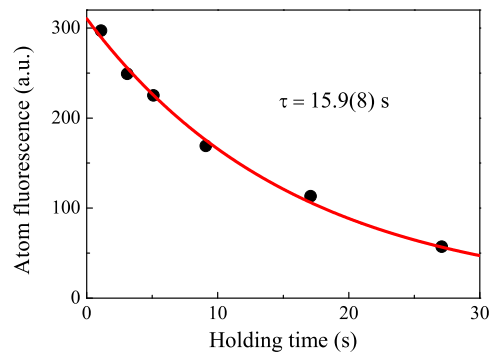}
 \caption{\label{fig:lifetime} (color online). Lifetime of Rb atoms in a magnetic quadrupole trap loaded from a MOT filled with the help of LIAD. Black dots are atom fluorescences measured using MOT recapture after different holding times in the magnetic trap (see text for detail). The red line is an exponential fitting to the measurement, which gives a $1/e$ lifetime of $\sim$ 16 s. The measurement is carried out one second after turning the UV light off. }
 \end{figure}

We have measured the magnetic trap lifetime by MOT recapturing. After different holding times, the atoms are released from the quadrupole trap by suddenly reducing the gradient from 160 G/cm to the value used for normal MOT operation. The MOT beams are then turned back on and the remaining atoms reveal themselves by fluorescences. As shown in Fig. \ref{fig:lifetime}, the 16 s trap lifetime coupled with the large atom number is a clear evidence of the LIAD technique's effectiveness.

Direct forced radio (RF) or microwave (MW) frequency evaporation in the quadrupole trap stops at a PSD of 10$^{-4}$ due to Majorana loss. To partially mitigate this loss, we superimpose a crossed ODT to the atoms together with the quadrupole field. The crossed ODT is produced by a 1070 nm, multi-frequency, linear polarized fiber laser(IPG Photonics). A 110 MHz acousto-optical modulator (Crystal Technology) is used for intensity stabilization and rapid trap switching off in less than 1 $\mu$s. We divide the laser power into two arms using a $\lambda$/2 waveplate and a polarizing beamsplitter cube. They intersect each other with an angle of $62^{\circ}$. The beam waists are 90 $\mu$m, while the vertical offset between the foci and the magnetic field zero is 150 $\mu$m. Only 4.5 W power is used in each beam, which produces a trap depth of 110 $\mu$K for $^{87}$Rb. They are ramped up in 200 ms after the quadrupole trap has reached its full strength and remain there during the MW evaporation. 

\begin{figure}[hbtp]
\includegraphics[width=0.85\linewidth]{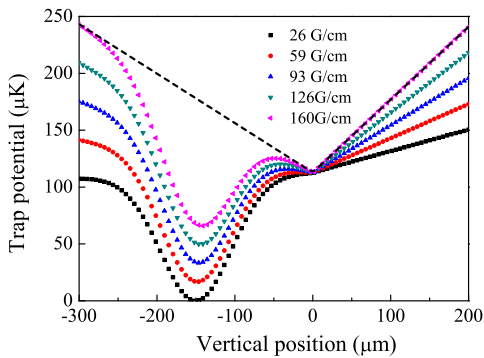}
\caption{\label{fig:potential} (color online). Variation of the hybrid trap potential along $y$ (gravity) direction during the magnetic quadrupole gradient ramping down. The black dashed line is the magnetic trap potential at 160 G/cm without the dipole trap. The dotted lines show the overall potential with different magnetic trap strengths ranging from gradients of 160 G/cm to 26 G/cm. The optical trap and magnetic trap centers are displaced by 150 $\mu$m from each other.}
\end{figure}

The effective potential for atoms in this hybrid trap is 
\begin{eqnarray}
U(r)=&&\frac{1}{2}\mu_B B^{'}\sqrt{\dfrac{x^2}{4}+y^2+\dfrac{z^2}{4}}+mgy\nonumber\\
&&-U_0 e^{-2(x^2+(y-y_0)^2)/w_0^2} + E_0
\label{eq:four}.
\end{eqnarray}
where $B'$ is the quadrupole field gradient along the vertical direction $\hat{y}$. $U_0$ is the optical trap depth at $y = y_0$  and $w_0$ is the beam waist. $E_0$ is the potential difference between the quadrupole trap center and the final trap minimum. $\mu$$_B$, $m$ and $g$ are the Bohr magneton, $^{87}$Rb atomic mass and the acceleration of gravity, respectively. As illustrated in Fig. \ref{fig:potential}, at 160 G/cm, there are two potential minima as a result of the displaced optical and magnetic trap centers. The potential difference between these two minima is $\sim$ 40 $\mu$K. This is comparable to the 20 $\mu$K cloud temperature where Majorana loss becomes severe. Thus this loss cannot be suppressed completely.

\subsection{Majorana loss}

\begin{figure}[hbtp]
\includegraphics[width=0.85\linewidth]{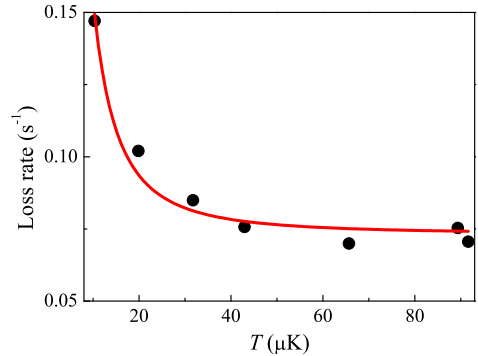}
 \caption{\label{fig:majorana} (color online). Majorana loss in the bare quadrupole trap. Black dots are measured atom loss rates at different temperatures obtained by MW evaporative cooling to different final frequencies. The red solid curve is a fit of the data with a loss model.(see text for detail).}
 \end{figure}

In an effort to quantify the Majorana loss in this setup, we evaporate the atoms to different temperatures by controlling the final evaporation cut frequency, and then measure the lifetimes to obtain loss rates. The measured data points are shown in Fig. \ref{fig:majorana}. Following ref.\cite{Petrich95}, the Majorana induced loss rate can be estimated as:
\begin{equation}
\Gamma_m=\chi\dfrac{\hbar}{m}(\dfrac{0.5 \mu_B B'}{k_B T})^2. 
\label{eq:AtomLoss}
\end{equation}
Here $\chi$ is a proportional constant, $T$ is the temperature, $\hbar$ is the Planck constant over 2$\pi$ and $k_B$ is the Boltzmann constant. 

To account for the background loss, we modify this equation as $\Gamma_L=aT^{-2}+\Gamma_b$ \cite{Heo11}, where $\Gamma_L$, and $\Gamma_b$ are the measured total loss rate and the background loss rate, respectively. The coefficient $a =\chi\dfrac{\hbar}{m}(\dfrac{0.5 \mu_B B'}{k_B})^2$, as a result of equation (\ref{eq:AtomLoss}). The fitting yields $a = 8.3(9)\mu K^2/s$ and $\Gamma_b = 0.071(4)$. Further analysis is not pursued due to the comparably large background loss rate, but the $T^{-2}$ dependence is already evident. 

For the full strength quadrupole trap, we have observed that Majorana loss is still severe and evaporative cooling below 20 $\mu$K is inefficient even with the displaced ODT beams. This is consistent with our understanding of the potential following Fig. \ref{fig:potential}.  The situation is marked improved after the magnetic gradient is reduced to 26 G/cm. With the ODT beam still at full power, the trap lifetime is measured to be 13(1) s for a 8 $\mu$K cloud. This matches with the background limited value well and suggests that the Majorana loss is fully suppressed.

\section{BEC production}
\subsection{Evaporation in the magnetic trap}
Evaporative cooling in magnetic trap starts right after the loading is finished. Condensates of similar numbers can be obtained with either RF or MW evaporation. Here only the current MW setup will be described. The MW signal driving the transition between $\left|1,-1\right\rangle$ and $\left|2,-2\right\rangle$ hyperfine levels is generated by doubling the  output of a signal generator (Anapico ASPIN6000-HC). After a 3 W amplifier (Minicircuit ZVE-3W-183), it is broadcast to the atoms by a microwave horn antenna. Thanks to the excellent initial PSD, this evaporation step can be completed within 6 seconds during which the frequency is swept from 6774 MHz to 6822 MHz. The sweep is divided into several segments with different slopes and powers. This is done empirically by optimizing the PSD obtained after each segment. During the whole procedure, a truncation factor of $\eta$ $\approx$ 6 is roughly maintained. After the MW evaporation, we typically end up with $2.5\times10^7$ atoms at temperatures of $\sim$ 29 $\mu$K and calculated densities of 10$^{12}$/cm$^3$. This corresponds to a PSD of $7\times10^{-5}$.    

\subsection{Transfer to the optical dipole trap}
After the MW evaporation in magnetic trap, the atoms are transfered into the ODT by ramping down the $B'$ linearly from 160 G/cm to 26 G/cm in 500 ms. This final $B'$ value is chosen to below 30.5 G/cm, which is the minimum gradient required for levitating Rb $\left|1,-1\right\rangle$ atoms against gravity. During this process, the confinement provided by the magnetic trap gradually decreases. The atoms sag down under the influence of gravity until they are trapped by the ODT. The trap potential deformation during the quadrupole trap decompression is shown in Fig.\ref{fig:potential}. At the final $B$ field gradient, the atoms occupy mainly the near bottom part of the potential and Majorana loss is suppressed by the high potential wall.

\begin{figure}[hbtp]
\includegraphics[width=0.85\linewidth]{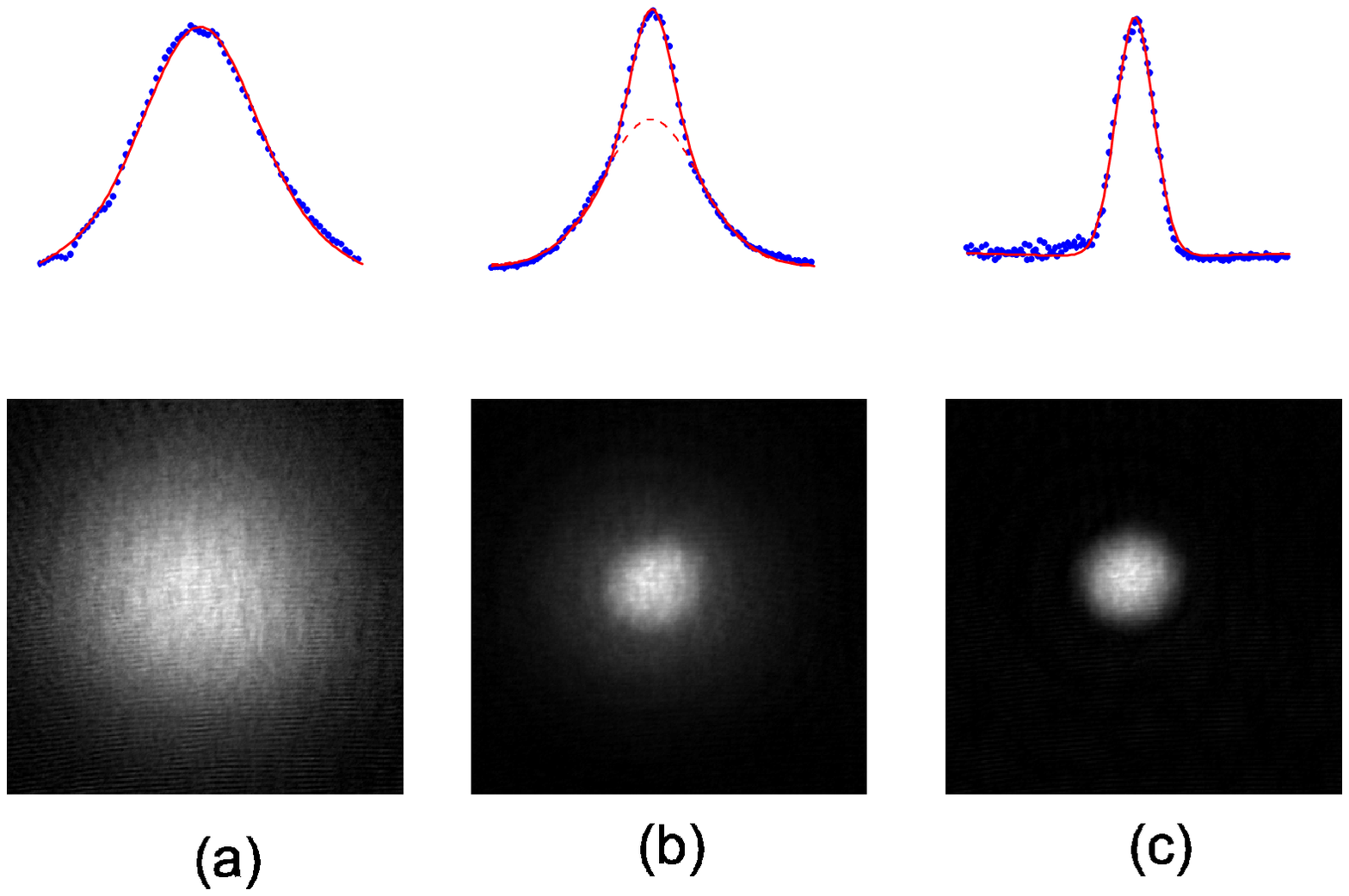}
\caption{\label{fig:BEC} (color online). Absorption images after 30 ms time of flight showing evidence of the BEC phase transition following evaporation in the ODT. Bottom panels: (a) thermal cloud at just above the transition temperature; (b) bimodal distribution; (c) a quasi-pure condensate with $10^5$ atoms. Field of view: 900 $\mu$m by 900 $\mu$m. Top panels: the integrated optical densities of corresponding images. Blue dots are experimental data, red solid lines are fittings to Gaussian (thermal atoms) or/and parabola (condensed atoms) functions. The red dashed line is the Gaussian fitting of thermal atoms in the bimodal phase. }
\end{figure}

The magnetic field gradient ramping speed is selected experimentally for best final condensate numbers. This rate is rather fast compared with the Lin et al.'s experiment \cite{Lin09} (2 seconds for the same gradient range). We suspect that this is due to the compromise between the larger background loss under our vacuum condition and the adiabaticity requirement. We also observe that no MW sweep is necessary in this step, while a factor of 40 increase in PSD is still observed because of continuous evaporation during the potential deformation. When the gradient reaches the final value, the atoms are loaded into the crossed ODT with temperatures of 14.6 $\mu$K, which is about one-eighth of the trap depth. Correspondingly, the PSD increases to $3\times10^{-3}$. The overall transfer efficiency from the quadrupole trap to the ODT is about $15\%$.

\begin{figure}[hbtp]
\includegraphics[width=0.85\linewidth]{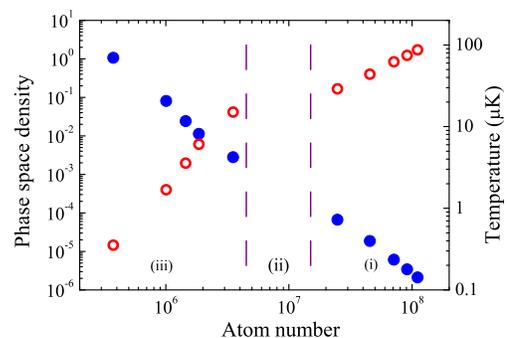}
\caption{\label{fig:trajectory}(color online). Evaporation trajectory showing the temperature(red circles) and peak phase-space density(blue dots) vs. atom number during the cooling process. Dashed vertical lines delimit three regions labeled(i),(ii)and (iii), which correspond to three evaporation steps. (i) forced MW evaporation in hybrid trap; (ii) loading into the dipole trap; (iii) evaporation in the optical dipole trap.}
\end{figure}

\subsection{Evaporation in the optical trap}
With $4\times10^6$ atoms of PSD $3\times10^{-3}$ in the crossed ODT, condensate production is straightforward. Forced evaporation is carried out by lowering the ODT power and thus the trap depth. We control the trap power following the scaling law $P(t) = P_i (1+t/\tau)^{-\beta}$, where $P_i$ is the initial power. Both $\tau$ and $\beta$ are determined experimentally for best final condensate number. According to the reference\cite{OHara01}, power ramping like this should result in optimal evaporation efficiency with fixed truncation factors. However, our case is complicated by thermal effect induced foci position shifts, mainly coming from the 3 mm thick Pyrex cell wall. These shifts will further loosen the trap in addition to the power reduction. Indeed, phase transition is already observed after lowering the power by a factor of 12. While following the scaling law, a reduction factor of $\sim$100 is needed. Judging from the final BEC number repeatability, we conclude that the thermal shift is reproducible to a great extent and thus further effort to minimize the shifts is not pursued. As shown in Fig. \ref{fig:trajectory}, the ODT evaporation efficiency defined as $\alpha = -\frac{log(PSD/PSD_0)}{log(N/N_0)}$ is 2.7, which is still quite high. Here PSD$_0$ and $N_0$ are the initial phase-space density and atom number, respectively.   

The ODT evaporation lasts for 3.5 s, which is also on the fast side. The phase transition happens at $T = 240$ nK with $2.7\times10^5$ atoms, signified by a bimodal distribution after 30 ms time-of-flight, as shown in Fig. \ref{fig:BEC}. Further evaporation leads to a quasi-pure condensate with $10^5$ atoms. The final trap frequencies are measured to be 2$\pi \times$($98,112, 61$)Hz, along $x$, $y$ and $z$ directions, respectively. 

We emphasize that the excellent initial PSD resulted short evaporation time is vital for producing condensates with relatively large numbers in our vacuum condition.  As shown in part (i) of Fig. \ref{fig:trajectory}, the PSD and temperature scale with the number as $N^{-2.17}$ and $N^{0.75}$ respectively in the magnetic trap evaporation. The evaporative is obviously less efficient compared with other experiments which typically have ten times better vacuum. In those cases, evaporation efficiencies close to $\alpha = 3$ have been achieved routinely \cite{Heo11,Lin09,McCarron11}. While the current evaporation efficiency is still high enough for reliable condensate production, with lower initial PSD, the evaporation has to be slowed down to allow enough thermalization. One-body number loss will limit the density gain in each evaporation steps and the evaporation efficiency could be further deteriorated. In the worst case, condensate production might even become impossible.

\section{Conclusion}

We have described a compact experimental setup capable of producing $^{87}$Rb Bose-Einstein condensates with more than $10^5$ atoms. Each experiment cycle takes about 30 seconds, contributed largely by the 20 s MOT loading time. The short evaporation time is made possible by the excellent initial PSD, and the high evaporation efficiency of the hybrid trap. Our success in condensate production with vacuum limited lifetime of 16 seconds further demonstrated the merit of the hybrid trap technique. Its combination with LIAD makes a nice shortcut to quantum gas productions. The current setup is limited by the MOT laser power. Using a high power tapered amplifier and larger beam size, higher initial number and shorter MOT loading times can be readily achieved. Larger condensates with even less evaporation time in similar setups are thus conceivable.

\begin{acknowledgments}
We thank Jie Ma for his contribution at the early stage of this experiment. We are grateful to Ruquan Wang and Shuai Chen for useful discussions and the loan of some essential optics. Our work was supported by Hong Kong RGC CUHK Grant No. 403111 and CUHK Direct Grant for Research No. 2060416. 
\end{acknowledgments}

\end{document}